\documentclass[preprint,12pt]{elsarticle}

\usepackage{amssymb}

\usepackage{amsmath,amsfonts}
\usepackage{mathrsfs}
\usepackage{algorithmic}
\usepackage{algorithm}
\usepackage{array}
\usepackage{textcomp}
\usepackage{url}
\usepackage{verbatim}
\usepackage{graphicx}
\usepackage{cite}
\usepackage{mathtools}
\usepackage{cases}
\usepackage[table,xcdraw]{xcolor}
\usepackage{upgreek}
\usepackage{caption}
\usepackage{listings}
\usepackage{fancyvrb}
\usepackage{soul}

\usepackage{tikz}
\usepackage{pgfplots,pgfplotstable}
\usetikzlibrary{babel} 
\usetikzlibrary{arrows,shapes,automata,backgrounds,petri,calc,fadings,3d}
\usetikzlibrary{decorations.markings,decorations.pathmorphing,decorations.pathreplacing}
\usepgfplotslibrary{groupplots,units,fillbetween,polar}
\usetikzlibrary{arrows.meta}
\tikzset{>=latex}
\usetikzlibrary{shapes.geometric, arrows}
\usetikzlibrary{arrows.meta}
\tikzset{>=latex}
\usetikzlibrary{calc}
\usepackage{makecell}%
\tikzstyle{startstop} = [rectangle, rounded corners, 
minimum width=3cm, 
minimum height=1cm,
text centered, 
draw=black, 
fill=red!30]
\pgfplotsset{compat=1.18}
\tikzstyle{io} = [trapezium, 
trapezium stretches=true, 
trapezium left angle=70, 
trapezium right angle=110, 
minimum width=3cm, 
minimum height=1cm, text centered, 
draw=black, fill=blue!30]

\tikzstyle{process} = [rectangle, 
minimum width=3cm, 
minimum height=1cm, 
text centered, 
text width=3cm, 
draw=black, 
fill=orange!30]

\tikzstyle{decision} = [diamond, 
minimum width=4cm, 
minimum height=2cm, 
text centered, 
draw=black, 
fill=green!30]
\tikzstyle{arrow} = [thick,->]

\definecolor{myazul}{rgb}{0,0.4431,0.7373}
\definecolor{myazul2}{rgb}{0,0.25,0.75}
\definecolor{mynaranja}{rgb}{0.8471,0.3216,0.0941}
\definecolor{myverde}{rgb}{0.47,0.67,0.19}
\definecolor{myverde2}{rgb}{0.0,0.5,0.0}
\definecolor{myvioleta}{rgb}{0.4,0.18,0.56}
\definecolor{myvioleta2}{rgb}{0.4,0.18,0.86}
\definecolor{myceleste}{rgb}{0.6,0.85,0.93}
\definecolor{myamarillo}{rgb}{0.93,0.88,0.17}
\definecolor{myrojo}{rgb}{0.84,0.04,0.00}
\definecolor{myrojo2}{rgb}{0.71,0 ,0}
\definecolor{mynegro}{rgb}{0.01,0.01,0.01}
\definecolor{backgroundColour}{rgb}{0.95,0.95,0.92}

\newcommand{\abs}[1]{\left|#1\right|}

\newcommand{\deff}{d_{\mathrm{eff}}}
\newcommand{\lcr}{L_{\mathrm{cr}}}

\newcommand{\derparvar}[2]{\frac{\partial #1}{\partial #2}}

\newcommand{\dk}{\Delta k}

\newcommand{\fourier}[1]{\mathcal{F}\left\lbrace #1 \right\rbrace}
\newcommand{\invfourier}[1]{\mathcal{F}^{-1}\left\lbrace #1 \right\rbrace}

\def\cpp{{C\nolinebreak[4]\hspace{-.05em}\raisebox{.4ex}{\tiny\bf ++}}}

\newcounter{bla}

\journal{Computer Physics Communications}

\begin{document}

\begin{frontmatter}



\title{CUDA-based focused Gaussian beams second-harmonic generation efficiency calculator}


\author[a]{A. D. Sanchez\corref{cor1}}
\cortext[cor1] {Corresponding author.\\\textit{E-mail address:} alfredo.sanchez@icfo.eu}
\author[b]{S. Chaitanya Kumar}
\author[a,c]{M. Ebrahim-Zadeh}

\address[a]{ICFO-Institut de Ciencies Fotoniques, Mediterranean Technology Park, 08860 Castelldefels, Barcelona, Spain.}
\address[b]{Tata Institute of Fundamental Research Hyderabad, 36/P Gopanpally, Hyderabad 500046, Telangana, India.}
\address[c]{Instituciò Catalana de Recerca i Estudis Avancats (ICREA), Passeig Lluis Companys 23, Barcelona 08010, Spain.}

\begin{abstract}

We present an object-oriented programming (OOP) CUDA-based package for fast and accurate simulation of second-harmonic generation (SHG) efficiency using focused Gaussian beams. The model includes linear as well as two-photon absorption that can ultimately lead to thermal lensing due to self-heating effects. Our approach speeds up calculations by nearly 40x (11x) without (with) temperature profiles with respect to an equivalent implementation using CPU. The package offers a valuable tool for experimental design and study of 3D field propagation in nonlinear three-wave interactions. It is useful for optimization of SHG-based experiments and mitigates undesired thermal effects, enabling improved oven designs and advanced device architectures, leading to stable, efficient high-power SHG.

\end{abstract}


\end{frontmatter}



{\bf PROGRAM SUMMARY/NEW VERSION PROGRAM SUMMARY}

\begin{small}
\noindent
{\em Program Title: \verb|cuSHG|}                                          \\
{\em CPC Library link to program files:} (to be added by Technical Editor) \\
{\em Developer's repository link: https://github.com/alfredos84/cuSHG} \\
{\em Code Ocean capsule:} (to be added by Technical Editor)\\
{\em Licensing provisions: MIT}  \\
{\em Programming language: \cpp, CUDA}                                   \\
{\em Supplementary material:}                                 \\
{\em Journal reference of previous version:}*                  \\
{\em Does the new version supersede the previous version?:}*   \\
{\em Reasons for the new version:*}\\
{\em Summary of revisions:}*\\
{\em Nature of problem: The problem that is solved in this work is that of second-harmonic generation (SHG) performance degradation in a nonlinear crystal with focused Gaussian beams due to thermal effects. By placing the nonlinear crystal in an oven that controls temperature, the package computes the involved electric fields along the medium. The implemented model includes the linear and nonlinear absorption which occasionally lead to self-heating effect, degrading the performance of the SHG.}\\
{\em Solution method: The coupled differential equations for three-wave interactions, which describe the field evolution along the crystal are solved using the well-known Split-Step Fourier method. The temperature profiles are estimated using the finite-elements method. The field evolution and thermal effects} are embedded in a self-consistent algorithm that sequentially and separately solves the electromagnetic and thermal problems until the system reaches the steady state. Due to the eventual computational demand that some problems may have, we chose to implement the coupled equations in the \cpp/CUDA programming language. This allows us to significantly speed up simulations, thanks to the computing power provided by a graphics processing unit (GPU) card. The output files obtained are the interacting electric fields and the temperature profile, which have to be analyzed during post-processing.\\

\end{small}

\section{Introduction}
\label{sec:intro}
\noindent

The propagation of focused Gaussian beams in second-order nonlinear optical processes, in particular second-harmonic-generation (SHG), has been widely studied theoretically and experimentally since the invention of the laser~\citep{kleinman1966second,bjorkholm1966,boyd1968parametric}. The process corresponds to the interaction of two identical photons of a given frequency in a non-centrosymmetric medium exhibiting $\chi^{(2)}$ polarization, leading to the generation of a new photon at the output at twice the input frequency. SHG is a fascinating optical phenomenon with major implications across a wide range of applications in science and technology. This process, well-known for its ability to efficiently convert optical frequencies, has found its importance in diverse fields, such as laser sources~\citep{samanta2010multicrystal}, imaging~\citep{keikhosravi2014second}, and material characterization~\citep{matlack2015review}, only to name a few. The ability to control and optimize SHG efficiency is of critical importance for the performance of various optical devices and laser systems and has major implications for their usability in applications.

In experimental implementation of the process, the precise calculation of SHG efficiency is crucial before embarking on costly and time-consuming tasks of material procurement and system construction. Prior knowledge of the expected output power or intensity allows researchers to make informed decisions about experimental parameters, crystal properties, and overall feasibility. However, accurately predicting SHG efficiency can be a complex task due to the multitude of factors involved, including beam characteristics, crystal properties, and, significantly, two-photon absorption effects~\citep{sabouri2013thermal}. The demand for a rapid and efficient algorithm to calculate SHG efficiency is undeniable, as it not only saves valuable time but also provides important insights into the experimental design process.

We present a numerical package, \verb|cuSHG|, that offers a robust and high-performance solution for predicting single-pass SHG efficiency in practical experimental scenarios. Our package is scripted in \cpp~and CUDA and simulates the interaction of focused Gaussian beams in a second-order dielectric nonlinear crystal under perfect phase-matching conditions, also including the thermal as well as the linear and nonlinear optical properties of the medium. Specifically, the combination of linear and nonlinear absorption of the fundamental as well as SH beams in the crystal results in various effects including thermal lensing, longitudinal and transverse thermal gradients, beam quality distortion, and long-term power degradation. Detailed understanding and analysis of these affects is pivotal to the design, optimization, and characterization of high-power SHG sources. Using \verb|cuSHG|, we are able to simulate the high-power single-pass SHG process, accurately visualizing the influence of thermal effects on the performance characteristics of the SHG source, including the absolute values of SHG power, where we find good agreement with the experimental results ~\citep{kumar2011high}. While commercial software such as COMSOL provide solutions for calculating light propagation in nonlinear media, including non-centrosymmetric crystals, our open-access code provides additional insight by simulating thermal effects within the crystal using a self-consistent model. By accounting for these intricacies, \verb|cuSHG| provides an invaluable tool for researchers, engineers, and scientists, enhancing their ability to optimize and realize efficient SHG, while minimizing detrimental thermal effects. The paper is structured as follows. In Section~\ref{sec:theory}, we present the theoretical framework of the model that \verb|cuSHG| implements. Section~\ref{sec:numimpl} describes the algorithms involved the packages. In Section~\ref{sec:softdesc}, the package structure is described. In section~\ref{sec:examples}, we show some practical examples of the \verb|cuSHG| features, before presenting the conclusions in Section~\ref{sec:conclusions}.

\section{Theoretical framework}
\label{sec:theory}
\noindent

\begin{figure}
    \centering
    \includegraphics[width=1.0\textwidth]{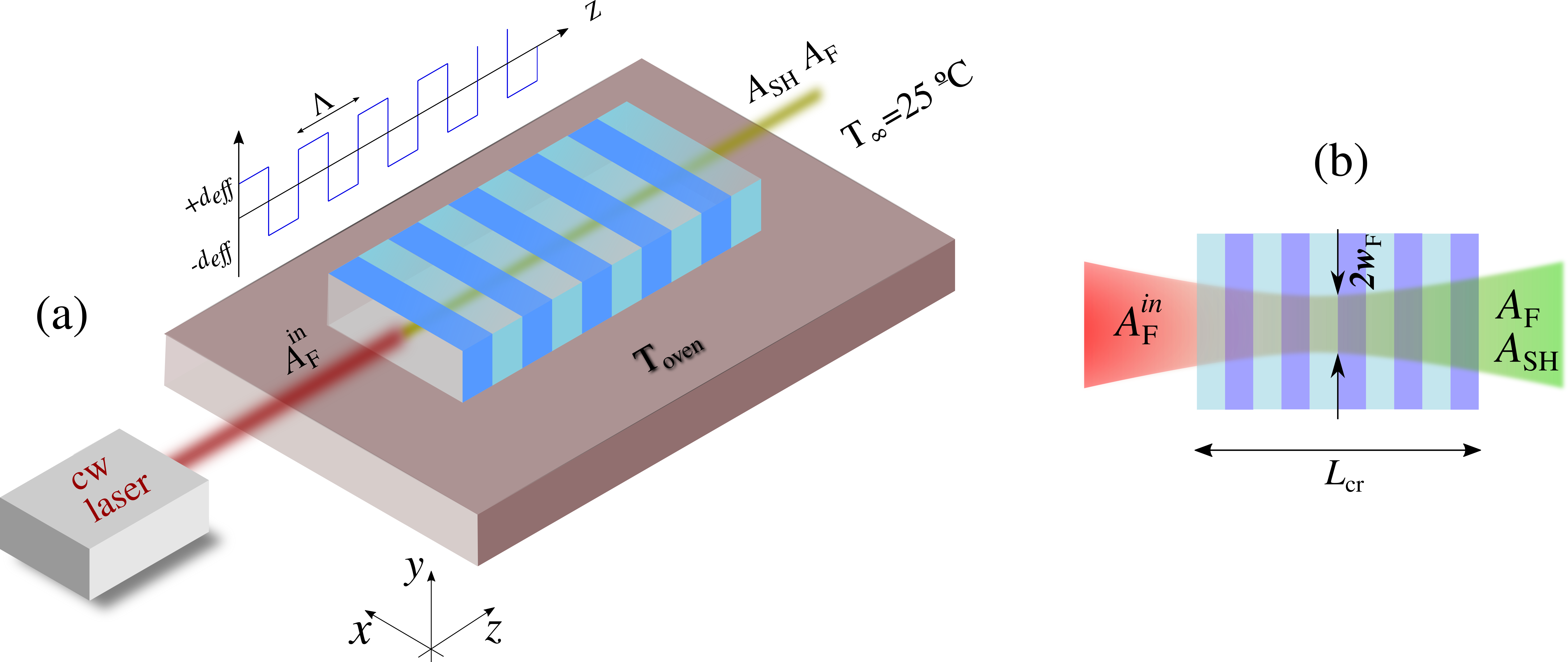}
    \caption{Schematic of the physical system that the package models. (a) practical setup; (b) focused Gaussian beam to produce SHG.}
    \label{fig:setup}
\end{figure}

In this section, we formulate the theoretical framework that models single-pass SHG of focused Gaussian beams in the presence of thermal effects. Since there is no unified model that describes the propagation of interacting electric fields in a nonlinear medium and how this affects the temperature of the medium, and vice versa, it is necessary to split the problem into two parts: electromagnetic propagation and thermal evolution. The electromagnetic part covers the solution of coupled differential equations, commonly known as \textit{coupled-wave equations} (CWEs). These equations are derived from Maxwell's equations with nonlinear polarization as the source of the electric field~\citep{shen1984principles}. On the other hand, the thermal evolution is solved using the well-known heat equation with internal source, in which the temperature profile is calculated at each point of the volume to be considered with experimental boundary conditions. The two parts are related as follows. The temperature profile changes the refractive index of the nonlinear medium locally, affecting its optical properties. In turn, the input as well as the generated electric field act as an internal source for the heat equation, resulting in a self-heating process. To solve the entire problem, we use a self-consistency algorithm in which the electric and temperature fields are successively calculated until a steady state is reached~\citep{seidel1997numerical}. Figure~\ref{fig:setup}(a) schematically shows the physical system that our package models. A continuous-wave (cw) laser is used as an input pump with electric field, $A_{\mathrm{F}}^{in}$, in a second-order nonlinear crystal to generate the second harmonic field, $A_{\mathrm{SH}}$. The chosen nonlinear crystal is periodically-poled stoichiometric lithium tantalate (MgO:sPPLT) with a grating period, $\Lambda$, to achieve SHG under temperature-tuned quasi-phase-matching~\citep{fejer1992quasi}. There are several possible configurations for temperature control of the crystal~\citep{sabouri2013thermal}. For our simulation, we use an open-top configuration, as shown schematically in Fig.~\ref{fig:setup}(a), where the oven at temperature, $T_{\mathrm{oven}}$, is in contact with the base of the crystal, with the rest of the structure exposed to ambient air at temperature, $T_{\mathrm{\infty}}$. Figure~\ref{fig:setup}(b) sketches a focused Gaussian beam of a given waist radius, $w_{\mathrm{F}}$, focused at the center of the nonlinear crystal of length, $\lcr$, to generate the second harmonic field.

\subsection{Electromagnetic problem}
The single-pass SHG is modeled using CWEs that well describe the  propagation of the interacting fields in a second-order nonlinear medium under the slowly varying envelope approximation~\citep{boyd2020nonlinear}. In our model, however, we also include thermal effects. We are interested in calculating the evolution of electric fields considering the effects of diffraction and losses, as well as the interaction mediated by second-order electric susceptibility within the nonlinear crystal. The CWEs for SHG are~\citep{new2011introduction}
\begin{numcases}{}
     \frac{\partial A_{\mathrm{F}}}{\partial z} = i\frac{2\pi \deff}{n_{\mathrm{F}}\lambda_{\mathrm{F}}} A_{\mathrm{SH}} A_{\mathrm{F}}^*e^{-i\phi(\Vec{r})} -\frac{i}{2k_{\mathrm{F}}}\nabla^2_{\perp} A_{\mathrm{F}} -\frac{1}{2}\left(\alpha_{\mathrm{F}}+\beta_{\mathrm{F}}\abs{A_{\mathrm{F}}}^2\right) A_{\mathrm{F}} \label{eq:cwep}\\
     \frac{\partial A_{\mathrm{SH}}}{\partial z} = i\frac{2\pi\deff}{n_{\mathrm{SH}}\lambda_{\mathrm{SH}}} A_{\mathrm{F}}^2e^{+i\phi(\Vec{r})} -\frac{i}{2k_{\mathrm{SH}}}\nabla^2_{\perp} A_{\mathrm{SH}} -\frac{1}{2} \left(\alpha_{\mathrm{SH}}+\beta_{\mathrm{SH}}\abs{A_{\mathrm{SH}}}^2\right) A_{\mathrm{SH}} \label{eq:cwes},
\end{numcases}
where $\Vec{r}=(x,y,z)$ is the position vector in the crystal, $z\in\left[0,\lcr\right]$ is the coordinate along the propagation direction, $\lcr$ is the crystal length, the subscripts $i=\mathrm{F},~\mathrm{SH}$ stand for \textit{fundamental (or pump)} and \textit{second harmonic (or signal)}, $A_i$ are the electric fields, $k_i=2\pi n_i/\lambda_i$ is the magnitude of the wave vector inside the crystal at the wavelength $\lambda_i$ with refractive index $n_i$, that may occasionally depend on temperature. The symbol $\nabla_{\perp}^2=\partial^2/\partial x^2+\partial^2/\partial y^2$ represents the transversal Laplacian along the transversal coordinates, $x\in\left[-L_x/2,L_x/2\right]$ and $y\in\left[-L_y/2,L_y/2\right]$, and $\alpha_i$ and $\beta_i$ denote the one- and two-photon absorption coefficients. Finally, $\omega_i$ is the corresponding angular frequency, $\deff$ is the effective nonlinear susceptibility, and $c$ is the speed of light. The accumulated phase, $\phi(\Vec{r})$, is
\begin{equation}\label{eq:integralpm}
    \phi(\Vec{r}) = \int_{0}^{z}\dk(x,y,z^{\prime},T)\,dz^{\prime}, 
\end{equation}
where 
\begin{equation}\label{eq:mismatch}
     \dk(T(\Vec{r})) = \frac{4\pi}{\lambda_{\mathrm{F}}}\left[(n_{\mathrm{SH}}(T(\Vec{r}))-n_{\mathrm{F}}(T(\Vec{r}))\right] -\frac{2\pi}{\Lambda(T(\Vec{r}))}
\end{equation}
is the mismatch factor that accounts for the efficiency of the process and depends on the local temperature in the crystal, and $\Lambda(T(\Vec{r}))$ is the grating period for periodically-poled nonlinear crystal~\citep{fejer1992quasi}, including its dependence on temperature that, in turn, accounts for the thermal expansion of the chosen crystal as
$$\Lambda(T(\Vec{r})) = \Lambda(25^{\circ}C)\left[1+a_{\Lambda}(T(\Vec{r})-25^{\circ}C)+b_{\Lambda}(T(\Vec{r})-25^{\circ}C)^2\right],$$
where $a_{\Lambda}=2.2\times 10^{-6}$~T$^{-1}$ and $b_{\Lambda}=-5.9\times 10^{-9}$~T$^{-2}$ for MgO:sPPLT crystal~\citep{nikogosyan2006nonlinear}.

In the seminal paper of theory of focused Gaussian beams, an expression for SHG conversion efficiency as a function of the most relevant experimental parameters has been derived~\citep{boyd1968parametric}. The initial Gaussian pump beam is described by the electric field
\begin{equation}\label{eq:gaussianbeam}
A_{\mathrm{F}}(\Vec{r}) = \frac{A_{\mathrm{F}0}}{1+i\tau}\exp\left[-\frac{x^2+y^2}{w_{\mathrm{F}}^2(1+i\tau)}+ik_{\mathrm{F}}z\right]
\end{equation}
with 
$$ \tau = \frac{z-f}{z_R},~z_{R}=w_{\mathrm{F}}^2k_{\mathrm{F}}/2 , $$
where the subscripts $j=x,y$ are the transversal coordinates, $A_{\mathrm{F}0}$ is the electric field strength, $f$ and $w_{\mathrm{F}}$ are the focal points and the beam waists, and $z_{R}$ is the Rayleigh range. Equation~\ref{eq:gaussianbeam} describes a Gaussian beam propagating along the $z$ direction. For a beam focused at the center of the crystal, $f = \lcr/2$, the initial electric field at the entrance of the nonlinear crystal is
\begin{equation}\label{eq:gaussianbeamc}
A_{\mathrm{F}}(x,y,z=0) = \frac{A_{\mathrm{F}0}}{1-i\xi}\exp\left[-\frac{x^2+y^2}{w_{\mathrm{F}}^2(1-i\xi)}\right],
\end{equation}
where the \textit{focusing parameter}, defined as $\xi=\lcr/2z_{R}$, is widely used in the literature~\citep{kumar2009high,sabouri2013thermal}. The focusing parameter is of crucial practical importance, as it determines the conversion efficiency in SHG experiments. We set the initial electric field at the SHG wavelength as Gaussian white noise with both random amplitude and phase.

\subsection{Thermal contribution}
Thermal effects are prevalent in high-power SHG experiments which involve focused Gaussian beams in the nonlinear crystal. Although finite absorption at the fundamental and SH wavelengths initially results in a Gaussian temperature distribution in the nonlinear crystal at low power levels, the change in the local refractive index mediated by the thermo-optic coefficient of the material significantly affects the phase-matching condition at high power levels. Often, the linear absorption at the SH wavelength is an order of magnitude higher than that of the fundamental~\citep{louchev2005thermal}, particularly in the green, causing a dominant effect on the resultant complicated temperature profile, eventually leading to catastrophic damage. Further, the ability to maintain a uniform temperature throughout the nonlinear crystal requires a suitable oven design~\citep{sabouri2013thermal} and efficient means of heat confinement/extraction. Hence, nonlinear crystals with negligible linear and nonlinear absorption and good thermal conductivity are desirable to minimize longitudinal and transverse thermal gradients, thereby enabling high-power SHG. In this context, a detailed understanding of the temperature profiles inside the nonlinear crystal at a given power level is critical for optimization of the SHG process, which can only be obtained by simulations (Eq.~\ref{eq:mismatch}).

The conventional method to calculate the temperature in a generic volume, as is the case of a nonlinear crystal, is by solving the heat equation with the appropriate boundary conditions and with the internal sources. The heat equation in the steady-state reads
\begin{equation}\label{eq:heatEqSS}
    k\nabla^2 T(\Vec{r}) + \dot{q}(\Vec{r})= 0,
\end{equation} 
where $T(\Vec{r})$ is the temperature field, $k$ is the thermal conductivity (assumed to be constant), and $\dot{q}(\Vec{r})$ is the internal heat source, defined in our specific problem as~\citep{louchev2005thermal}
\begin{equation}
    \dot{q}(\Vec{r}) = \sum_{i\in \{\mathrm{F},\mathrm{SH}\}} \alpha_i I_i(\Vec{r}) + \beta_i I_i^2(\Vec{r}),
\end{equation}
where the intensity of the involved electric fields is defined as $I = \epsilon_0cn\abs{A}^2/2$, with $\epsilon_0$ the vacuum permittivity.

The nonlinear crystal is usually subjected to boundary conditions set by the surrounding air as well as an oven or Peltier cell that sets the temperature of one or more faces (see Fig~\ref{fig:setup}(a)). This fixes the temperature of the crystal to satisfy the phase-matching condition in Eq.~\ref{eq:mismatch}. The faces with a constant temperature satisfy the Dirichlet boundary condition
\begin{equation}\label{eq:toven}
    T(\Vec{r}_{\mathrm{f}})=T_{\mathrm{oven}},    
\end{equation}
where the subscript 'f' stands for all the points in the \textit{face}. We consider free convection for the faces in contact with the air. Therefore, the von Neumann boundary condition is given by
\begin{equation}\label{eq:convection}
    -k\left.\frac{\partial T}{\partial n}\right|_{\mathrm{f}} = h (T_{\mathrm{f}}-T_{\infty}),
\end{equation}
where $h$ is the heat transfer coefficient and $T_{\infty}$ is the surrounding temperature.

\section{Numerical implementation}
\label{sec:numimpl}
\noindent

\subsubsection{Split-step Fourier method}
\label{sec:ssfm}
\noindent

The crystal of length, $\lcr$, is discretized along the $z-$direction into steps of length, $dz$. In every step, the Split-step Fourier method (SSFM) simultaneously solves the linear and nonlinear effects. The linear part is solved in the spatial domain, and the nonlinear part is solved in the time domain using a four-order Runge-Kutta method. This sequence is repeated throughout the length of the crystal. Depending on the implementation, this algorithm exhibits an error, $\mathcal{O}(dz^2)$ or $\mathcal{O}(dz^3)$~\citep{sanchez2024cuda}. This is sequentially solved along the entire crystal and requires many operations with complex vectors as well as 2D discrete Fourier transforms (DFTs) during the simulation. 
Equations~\ref{eq:cwep} and~\ref{eq:cwes} can be written in the matrix form as 
\begin{equation}\label{eq:CWEsmatrix}
    \frac{\partial}{\partial z} 
    \begin{pmatrix}
    A_{\mathrm{F}}\\
    A_{\mathrm{SH}}
    \end{pmatrix} = \left(
    \underbrace{\begin{pmatrix}
    \hat{L}_{\mathrm{F}} & 0 \\
    0 & \hat{L}_{\mathrm{SH}}
    \end{pmatrix}}_{\text{Linear operator}\hat{L}}   
    +
    \underbrace{\begin{pmatrix}
    -\frac{1}{2}\beta_{\mathrm{F}}\abs{A_{\mathrm{F}}}^2 & i K_{\mathrm{F}}A^*_{\mathrm{F}} \\
     i K_{\mathrm{SH}}A_{\mathrm{F}} & -\frac{1}{2}\beta_{\mathrm{SH}}\abs{A_{\mathrm{SH}}}^2
    \end{pmatrix}}_{\text{Nonlinear operator~}\hat{N}}   
    \right)\cdot
    \begin{pmatrix}
    A_{\mathrm{F}}\\
    A_{\mathrm{SH}}
    \end{pmatrix},
\end{equation}
where 
$$\hat{L}_j=-\frac{i}{2k_j}\nabla^2_{\perp} A_j - \frac{\alpha_j}{2}$$
is the linear operator at the corresponding wavelength, with $j\in\{\mathrm{F},\mathrm{SH}\}$, and $K_{\mathrm{F}}= 2\pi\deff e^{-i\dk z}/n_{\mathrm{F}}\lambda_{\mathrm{F}}$ and $K_{\mathrm{SH}}= 2\pi\deff e^{+i\dk z}/n_{\mathrm{SH}}\lambda_{\mathrm{SH}}$. Equation~\ref{eq:CWEsmatrix} is then reduced to its vectorial form as
\begin{equation}
    \derparvar{\Vec{A}}{z} = \left( \hat{L} + \hat{N} \right) \Vec{A}
\end{equation}
with a symbolic solution given by
\begin{equation}\label{eq:evol}
    \Vec{A}(z+dz) = e^{\left( \hat{L} + \hat{N} \right)dz} \Vec{A}(z).
\end{equation}

Since the operators, $\hat{L}$ and $\hat{N}$, in general do not commute, the approximation, $e^{\left( \hat{L} + \hat{N} \right)dz} \approx e^{ \hat{L}dz} e^{\hat{N} dz}$, that yields an error, $\mathcal{O}(dz^2)$, is often used. 

However, in this work, we implement a more accurate expression~\citep{agrawal2000nonlinear}
\begin{equation}\label{eq:expon}
    e^{\left( \hat{L} + \hat{N} \right)dz} \approx e^{ \hat{N}\frac{dz}{2}}e^{ \hat{L}dz} e^{ \hat{N}\frac{dz}{2}},
\end{equation}
with an error of $\mathcal{O}(dz^3)$. In this scheme, every step is solved by computing the nonlinear term in the first half-step, $dz/2$. After one Fourier transform, the linear term is computed in the entire step, $dz$. Finally, the nonlinear term is again computed in the second half-step, $dz/2$. This sequence, $\hat{N}/2-\hat{L}-\hat{N}/2$, is equivalent to its counterpart, $\hat{L}/2-\hat{N}-\hat{L}/2$, since both lead to the same solution. By inserting Eq.~\ref{eq:expon} in Eq.~\ref{eq:evol}, and solving the linear part in the frequency domain, the field evolution reads
\begin{equation}
    \Vec{A}(z+dz) \approx e^{ \hat{N}\frac{dz}{2}} \invfourier{ e^{ \hat{L}dz} \fourier{e^{ \hat{N}\frac{dz}{2}}\Vec{A}(z)}} ,
\end{equation}
where $\fourier{\cdot}$ stands for the 2D-Fourier transform.~\ref{asec:diffraction} provides a more detailed description of the diffractive term.

\subsection{Finite differences}
\label{sec:findiff}

In order to solve Eq.~\ref{eq:heatEqSS}, we implement a standard finite difference scheme using the same grid as that in the case of SSFM. This implementation uses second-order derivatives and yields an accuracy error of $\mathcal{O}(dx^2+dy^2+dz^2)$. The temperature in the grid inner nodes as a function of the spatial coordinates and the internal source reads~\citep{holman2010heat}
\begin{equation}\label{eq:heatEqSSdisc}
\begin{split}
     \frac{T_{m+1,n,l}-2T_{m,n,l}+T_{m-1,n,l}}{\Delta x^2}+\frac{T_{m,n+1,l}-2T_{m,n,l}+T_{m,n-1,l}}{\Delta y^2}+ \\
     \frac{T_{m,n,l+1}-2T_{m,n,l}+T_{m,n,l-1}}{\Delta z^2} = - \frac{\dot{q}_{m,n,l}}{k},
\end{split}
\end{equation}
where the subscripts $m \in \left[0,N_x\right]$, $n \in \left[0, N_y\right]$, $l \in \left[0,N_z\right]$ are the indices in the chosen 3D grid for the spatial coordinates, $x,~y,~z$, respectively, with $N_i$ the number of nodes for each dimension. The nodes in the edges and faces are similarly calculated using the boundary conditions in Eq.~\ref{eq:toven} and \ref{eq:convection}.

\section{Package description}
\label{sec:softdesc}
\noindent

As mentioned in Section~\ref{sec:intro}, \verb|cuSHG| is scripted in the \cpp/CUDA programming language, since both the electromagnetic and thermal evolution exhibit a high degree of parallelism. The package was tested on a Linux system and its functionality depends upon \verb|cuda-toolkits|~\citep{NVT}. Users who work with a Windows system can also use the package by installing the proper drivers following the instructions provided in Ref.~\citep{NVT}.
The package contains a main file, a folder with header files, and a bash file that allows the user to compile and execute the package by varying the relevant simulation parameters. In our code, the electric and temperature fields, phase-matching function and the SSFM, are objects belonging to their corresponding classes that interact through their methods throughout the simulation.

The package contains ten header files in the folder, \verb|headers|, which can be either modified or adapted to the user specific applications. 
\begin{itemize}
    \item \verb|Libraries.h| contains the list of the required libraries related to \cpp~and CUDA programming.
    \item \verb|PackageLibraries.h| contains the set of the package libraries listed in the next items.
    \item \verb|Common.h| contains functions necessary to check other functions executed on the GPU.
    \item \verb|Operators.h| contains overloaded operators ($+,-,*,/$) to perform operations with real and complex numbers.
    \item \verb|Files.h| contains functions useful to save real or complex vectors, matrices and tensors into a \verb|.dat| file.
    \item \verb|Crystal.h| contains the Sellmeier equations for the predefined nonlinear crystals, as well as other relevant physical quantities set as a global constants. We choose MgO:sPPLT nonlinear crystal for our model~\citep{bruner2003temperature,nikogosyan2006nonlinear}. For other crystals, the user should modify this file accordingly with the new parameters.
    \item \verb|Efields.h| contains the definition of the \verb|class Efields| and its methods.  
    \item \verb|Tfield.h| contains the definition of the \verb|class Tfield| and its methods. In this library, the finite-different element method in the method \verb|upDate()|, which updates the crystal temperature in each iteration step, can be found.
    \item \verb|PhaseMatching.h| contains the definition of the \verb|class PhaseMatching| and its methods.
    \item \verb|Solver.h| contains the definition of the \verb|class Solver| and its methods. This library contains the SSFM routines.
\end{itemize}

The main file, called \verb|cuSHG.cu|, is divided into four short parts:
\begin{enumerate}
    \item \textit{Set GPU and timing}: Sets the intended GPU and starts the simulation timing.
    \item \textit{Set input parameters}: Defines the 8 simulation parameters required to run the code, namely, pump power, focal point, beam waist, crystal temperature, environmental temperature, oven temperature, and two Boolean variables to control the \verb|.dat| output files.
    We define the single-precision data types, \verb|real_t| and \verb|complex_t| (\verb|float| and \verb|cufftComplex|, respectively), that are needed to define scalars and vectors.
    \item \textit{Model execution}: executes the theoretical model described in Section~\ref{sec:theory}. The command line \verb|Solver *solver = new Solver;| created an instance of the object \verb|Solver|. Then, the model is executed using the method \verb|solver->run(<parameters>)| that receives the parameters set previously. After execution, command line \verb|delete solver| is used to destroy the object and free memory.
    \item \textit{Reset the GPU and finish simulation timing}: Finishes and returns the simulation runtime.
\end{enumerate}
Listing~\ref{lst:mainfile} briefly shows the main file with relevant parts. 
\begin{lstlisting}[caption={Main file structure.},label={lst:mainfile},language=C++, basicstyle=\ttfamily,
  showstringspaces=false,
  commentstyle=\color{gray!50!black},
  keywordstyle=\color{blue},
  basicstyle=\ttfamily\footnotesize,]
#include "headers/Libraries.h"	// Required libs.
using real_t = float;           // Datatypes for real 
using complex_t = cufftComplex; // and complex numbers
// Set global constants...
#include "headers/PackageLibraries.h" // Package libs.

int main(int argc, char *argv[])
{
    // 1. Set GPU and timing	
    // code  -------------------------------------
    // 2. Set input parameters
    // code  -------------------------------------
    // 3. Model execution
    Solver *solver = new Solver;	
    solver->run( PARAMETERS );
    delete solver;
    //--------------------------------------------
    // 4. Reset GPU and finish simulation timing
    cudaDeviceReset();
    TimingCode(iElaps); // print time
    //--------------------------------------------
    return 0;	
}  
\end{lstlisting}

The user can also find the function description in the source code and the full code overview in the \verb|README.md| file in the corresponding repository~\citep{repo}.

\subsection{Compilation and execution}
\label{sec:softfunctionalities}
\noindent
Before running the code, it is necessary to compile the package and obtain an executable file. To do this, we execute the bash file, \verb|cuSHG.sh|, included in the package, which in turn contains the compilation and the execution command lines. Before describing the command line to compile the package, it is important to note that the code can be executed in two ways:
\begin{enumerate}
    \item CWEs only: in simulations in which the thermal effects are ignored, the package can be compiled without temperature field calculation. 
    \item CWEs+Temperature: this compilation includes the above and ensures the execution of the model presented in Section~\ref{sec:theory}.
\end{enumerate}

The compilation command line is shown in Listing~\ref{lst:compilation}, where the compiler, \verb|nvcc|, is invoked to compile the file, \verb|cuSHG.cu|. We incorporate the preprocessor variable, \verb|-DTHERMAL|, to distinguish the two compilation modes previously described.

\begin{lstlisting}[caption={Compilation},label={lst:compilation},language=bash, basicstyle=\ttfamily,
  showstringspaces=false,
  commentstyle=\color{red},
  keywordstyle=\color{blue},
  basicstyle=\ttfamily\footnotesize,]
# 1. Compilation excludes thermal calculation
nvcc cuSHG.cu --gpu-architecture=sm_75
    -lcufftw -lcufft -o cuSHG
# ...OR...
# 2. Compilation includes thermal calculation
nvcc cuSHG.cu -DTHERMAL --gpu-architecture=sm_75
    -lcufftw -lcufft -o cuSHG
# Execution with the passed arguments
./cuSHG $<SET_OF_6_ARGUMENTS_TO_PASS>        
\end{lstlisting}

As can be seen, there are some extra flags required for the compilation. The flag, \verb|--gpu-architecture=sm_75|, tells the compiler to use the specific GPU architecture. It is important to check the proper value for this flag according to the used GPU card. The flags, \verb|-lcufftw| and \verb|-lcufft|, are used for the Fourier transforms performed by \verb|CUDA|. 

After successfully compiling, the next step is to run the code. In the file, \verb|cuSHG.sh|, there are some variables that will be passed as an argument to the main file,  \verb|cuSHG.cu|. This, of course, can be modified by users who prefer just to set the variables values in the main file. However, this may be useful when users need to systematically vary a physical quantity, e.g. pumping level, beam waist, oven temperature, etc. Once the execution is finished, the output files are moved to a specific folder created by the file, \verb|cuSHG.sh|. The name of the created folder is related to the simulation parameters, but this can be changed according to the user requirements.

\subsection{Algorithmic flowchart}

\begin{figure}[htbp]
\centering
\begin{tikzpicture}[node distance=2cm,thick,scale=0.7, every node/.style={transform shape},->]
\draw[black, dashed, thick, fill=black!10!white] (-2.5,-5) rectangle (8,-20);
\node[font=\normalsize] at (5.5,-5.5) {\textbf{Computation on GPU}};
\node (start) [startstop] {Start};
\node (in1) [io, below of=start] {Set parameters};
\node (togpu) [io, below of=in1] {Set matrices in GPU};
\node (pro1) [process, below of=togpu] {Set $\dot{q}(\Vec{r})=0$};
\node (pro2) [process, below of=pro1] {Compute $T(\Vec{r})$};
\node (pro2p) [process, below of=pro2] {Compute $\dk(T(\Vec{r}))$};
\node (pro3) [process, below of=pro2p] {\makecell[c]{Compute CWEs\\$A_{\mathrm{F}}(\Vec{r}),~A_{\mathrm{SH}}(\Vec{r})$}};
\node (pro4) [process, below of=pro3] {Compute $\dot{q}(\Vec{r})$};
\node (dec1) [decision, below of=pro4, yshift=-1.5cm] {\makecell[c]{Solution\\self-consistent}};
\node (pro5) [process, right of=dec1, xshift=3cm] {\makecell[c]{Next iteration\\using last $\dot{q}(\Vec{r})$}};
\node (tocpu) [io, below of=dec1, yshift=-2.5cm] {\makecell[c]{Copy data GPU$\rightarrow$CPU\\Save data to files}};
\node (stop) [startstop, below of=tocpu] {Stop};
\draw [arrow] (start) -- (in1);
\draw [arrow] (in1) -- (togpu);
\draw [arrow] (togpu) -- (pro1);
\draw [arrow] (pro1) -- (pro2);
\draw [arrow] (pro2) -- (pro2p);
\draw [arrow] (pro2p) -- (pro3);
\draw [arrow] (pro3) -- (pro4);
\draw [arrow] (pro4) -- (dec1);
\draw [arrow] (dec1) -- node[anchor=east, above] {no} (pro5);
\draw [arrow] (dec1) -- node[anchor=south, right] {yes} (tocpu);
\draw [arrow] (tocpu) -- (stop);
\draw [arrow] (pro5) |- (pro2);

\end{tikzpicture}
\caption{Algorithm flowchart adapted from Ref.~\citep{seidel1997numerical}.}
\label{fig:flowchart}
\end{figure}
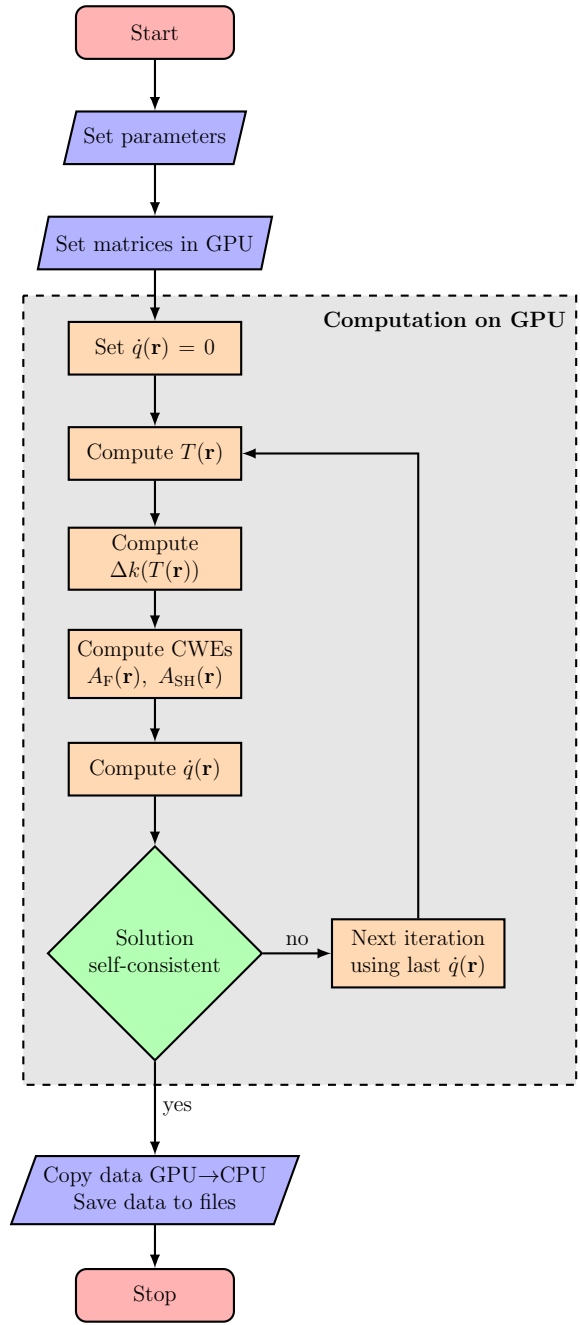

As mentioned previously, there is no unified model capable of calculating the involved electric and temperature fields simultaneously. The strategy to tackle this problem is to use an iterative model to obtain a self-consistent solution. Figure~\ref{fig:flowchart} shows the flowchart of the implementation adopted by our package.

\subsection{Package performance}
\label{sec:performance}
\noindent

The operations performed in our package are essentially sums, products, and discrete 2D-Fourier transforms (2DFT) of real- and complex-valued matrices. The use of a GPU is justified for a total number of NZ matrices with a size of $N$=NX$\times$NY used in our simulations. Here, NZ is the number of slices into which the crystal is divided along the $z$-direction, whilst NX and NY are the number of points in the transversal directions. The 2DFT on CPU were calculated with the widely used FFTW library, while on GPU, 2DFT were carried out using \verb|cuFFT|, the CUDA library for computing Fourier transforms. Both 2DFT implementations have an order of convergence, $\mathcal{O}\left(n\log(n)\right)$, with $N$ the involved number of points~\citep{frigo2005design, cuFFT}. On the other hand, the rest of operations of sums and products have an order of convergence, $\mathcal{O}(n)$. The global algorithm has a convergence order dominated by the 2DFT. 

\begin{figure}[hbtp]
    \centering
    \includegraphics[width=0.7\linewidth]{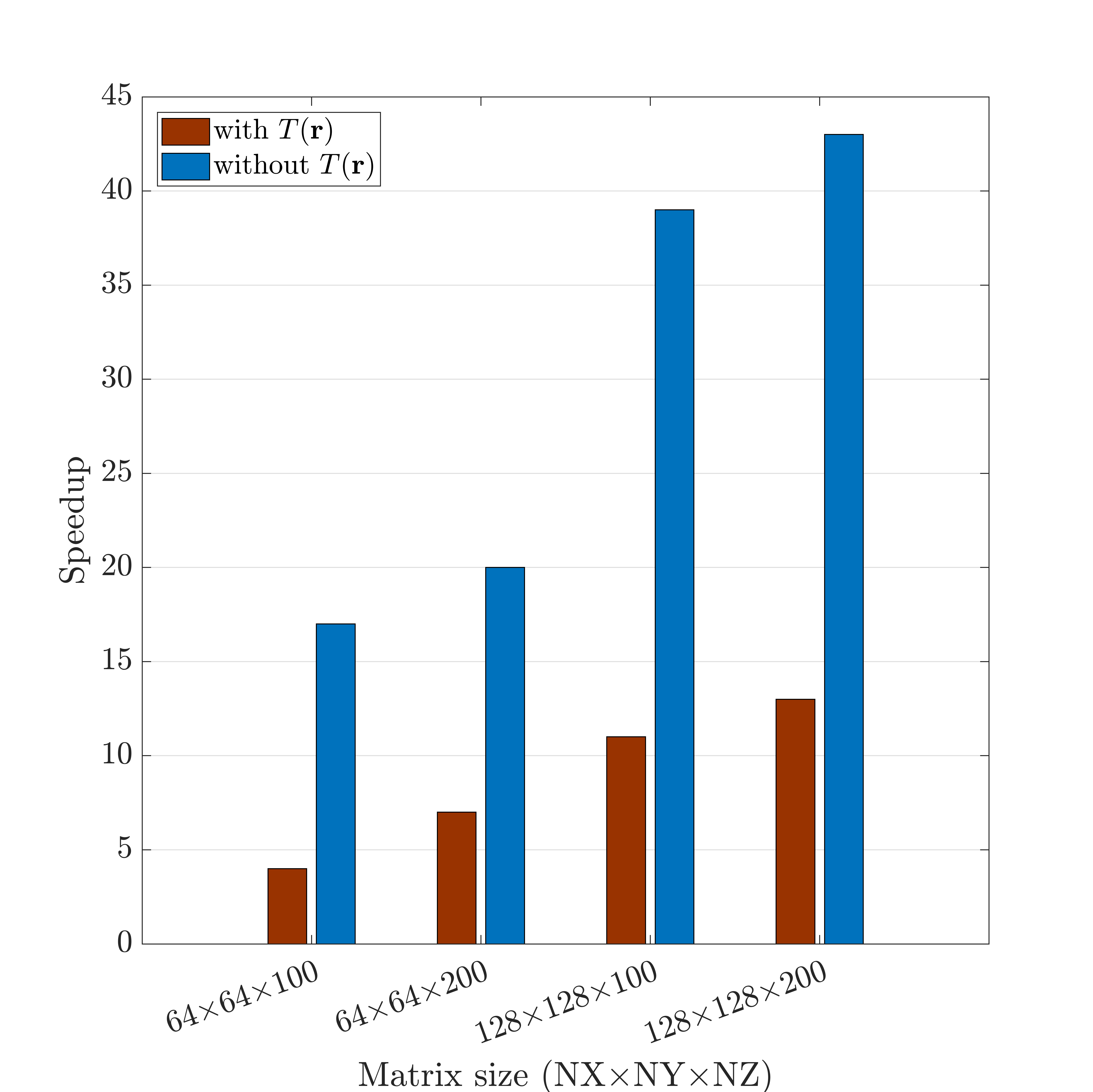}
    \caption{Speedup measurement for different matrix sizes.}
    \label{fig:performance}
\end{figure}

We compare the execution time of our model with an equivalent code scripted in \cpp~to be only executed in CPU. The ratio of the time for a given calculation performed on CPU to that obtained in GPU is called \textit{speedup}, and is a way to measure the performance of the GPU scripted algorithm. We measure the speedup of our package for typical values of NX=NY and NZ that are used in simulations. 
Figure~\ref{fig:performance} shows the obtained speedup using a desktop computer with a microprocessor Intel(R) Core(TM) i7-9700 CPU @ 3.00GHz and a GPU NVIDIA GeForce GTX 1650. 
As can be seen, the larger the size of the matrices, the greater the acceleration obtained. The speedup when calculating temperature profiles is minor and may be improved by implementing, e.g., a 2.5D stencil that uses the shared memory of the GPU~\citep{krotkiewski2013efficient}.

\section{Illustrative Examples}
\label{sec:examples}
\noindent

In the following examples, we show how to use the package in order to obtain the SHG conversion efficiency under different conditions. The first illustrative example is useful to test the correct functioning of the package by comparing the software outcome with a theoretical formula for SHG efficiency. The second example shows the full model execution to compute the thermal profile inside the crystal that results from the nonlinear interaction of the involved electric fields.

\subsection{Example 1: contrasting theoretical and numerical results}
\label{sec:example1}
\noindent

The calculation of SHG conversion efficiency using the theoretical model presented in Section~\ref{sec:ssfm} has an analytical solution under the specific conditions defined by Boyd and Kleinman (BK) in their seminal paper~\citep{boyd1968parametric}. In cases where pump depletion is considered, in addition to linear and nonlinear absorption, numerical simulations are required for efficiency calculation. Furthermore, when including thermal effects, there is no choice but to resort to numerical solutions. The BK-efficiency formula reads
\begin{equation}\label{eq:bkefficiency}
    \eta_{\mathrm{BK}} = \frac{P_{\mathrm{SH}}}{P_{\mathrm{F}}} = \frac{16\pi^2 \deff^2 P_{\mathrm{F}} \lcr}{\epsilon_0 c n_{\mathrm{F}} n_{\mathrm{SH}} \lambda_{\mathrm{F}}^3} h(\xi,\dk),
\end{equation}
where the function, $h(\xi,\dk)$, accounts for the phase-matching and the focusing parameter contributions. 

\begin{figure*}
    \centering
    \includegraphics[width=1.0\linewidth]{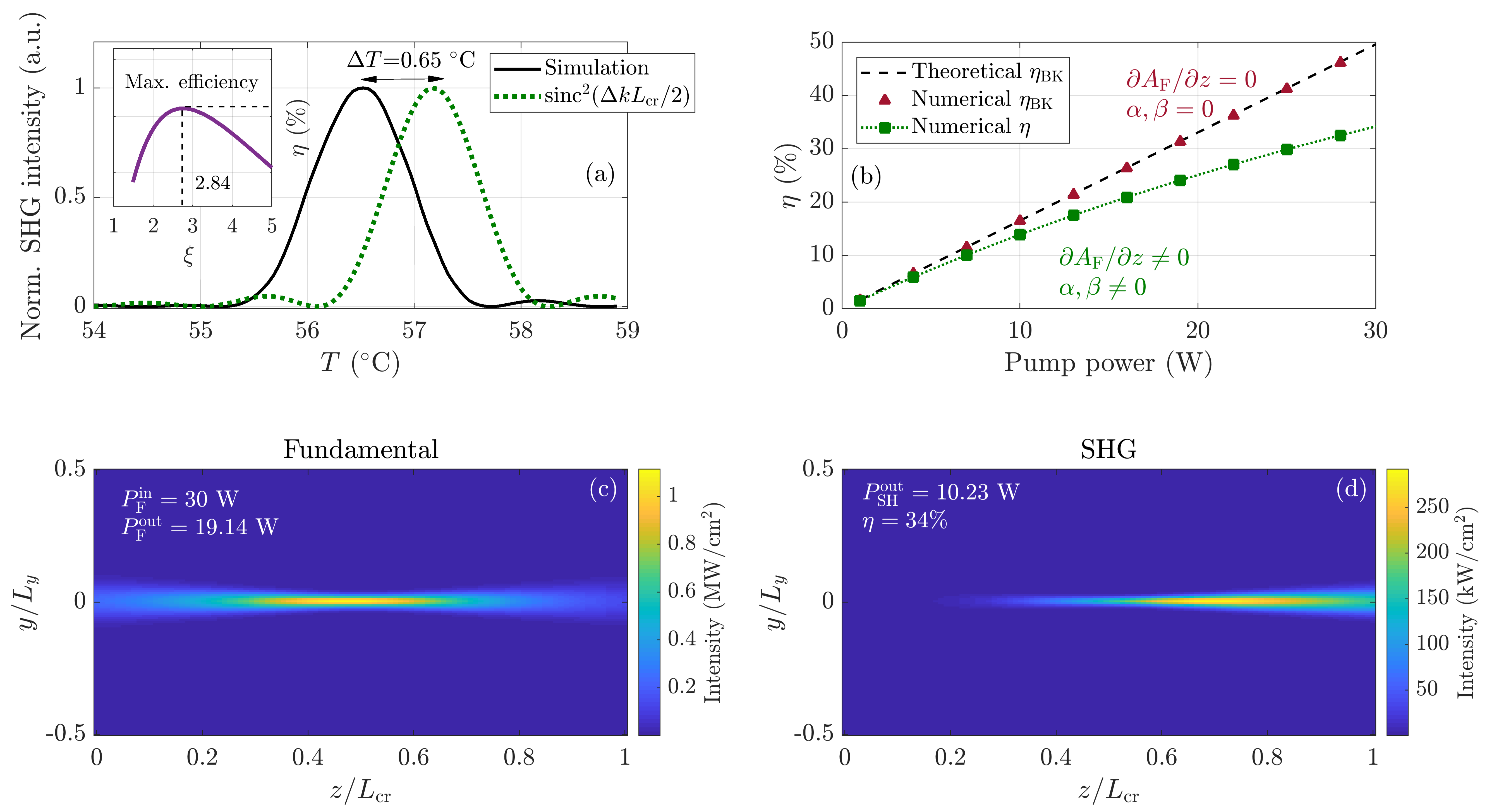}
    \caption{SHG efficiency calculation. Panel (a): Calculated and simulated temperature bandwidth. The inset shows that the maximum efficiency reached at $\xi=2.84$. Panel (b): BK efficiency, $\eta_{\mathrm{BK}}$, and the deviation including absorption and pump depletion. Panels (c,d): the pump electric field is focused at the center of the nonlinear crystal with a power of 30~W and a focusing parameter of $\xi=2.84$.}
    \label{fig:ex1}
\end{figure*}

Figure~\ref{fig:ex1}(a) shows the temperature bandwidth calculated from the phase matching condition for a pump power of 1~W and a pump beam waist of 40~$\upmu$m (green-dotted curve), and its comparison with the simulation results (solid-black curve). As expected, the simulated phase-matching temperature shifts with respect to theoretical calculation, since the focused Gaussian beams accumulate additional phase during the propagation~\citep{sabouri2013thermal}. The inset shows the calculated efficiency as a function of the focusing parameter, $\xi$, exhibiting a maximum value at $\xi=2.84$, in perfect agreement with the BK theory~\citep{boyd1968parametric}. Figure~\ref{fig:ex1}(b) shows the comparison of our numerical simulations with Eq.~\ref{eq:bkefficiency} ($\eta_{\mathrm{BK}}$), as well as the deviation when pump depletion, linear and nonlinear absorption are included (green squares). The undepleted pump condition, $\partial A_{\mathrm{F}}/\partial z = 0$, is achieved by setting the variable, \verb|dPump[IDX] = make_cuComplex(0.0f, 0.0f)|, to zero in the function, \verb|dAdz()|, placed in the header file \verb|Solver.h|. Figure~\ref{fig:ex1}(c),(d) show a cut in the $yz$-plane of the focused Gaussian beam evolution along the crystal for the fundamental and SH electric fields for a pump beam waist of $w_{\mathrm{F}}=29~\upmu$m and an oven temperature of $T=56.5~^{\circ}$C. 
The compilation of this example is detailed in Listing~\ref{lst:compilation}, case \verb|#1|, where the flag \verb|-DTHERMAL| is omitted. 

\subsection{Example 2: SHG efficiency including thermal effects}
\label{sec:example2}
\begin{figure*}
    \centering
    \includegraphics[width=1.0\linewidth]{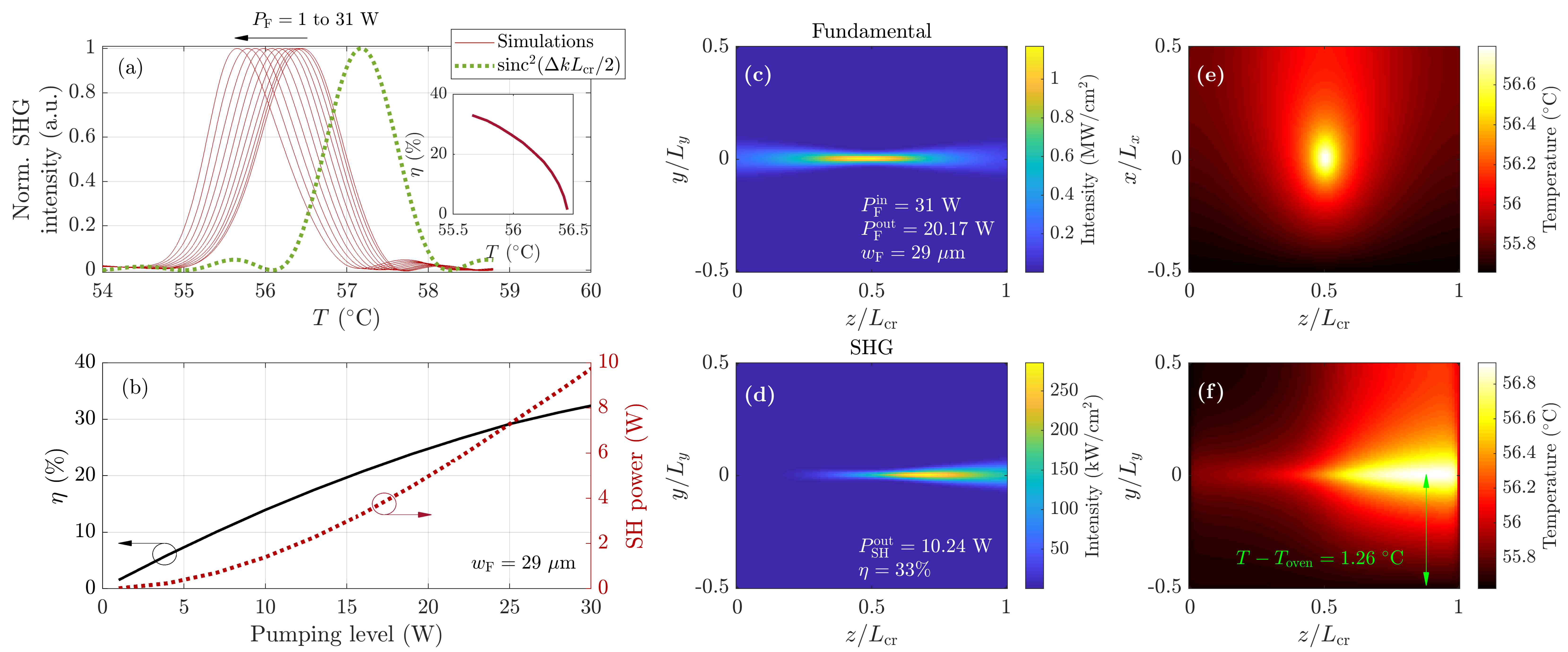}
    \caption{Simulations including thermal effects. Panel (a) Temperature bandwidth for different pumping levels.  The inset shows the computed maximum efficiency (peaks in the main plot) as a function of the temperature. Panel (b): Efficiency and Signal output power as a function of the pumping level. Panels (c): $yz$-plane cut for the pump intensity, measured in MW/cm$^2$. Panel (d) $yz$-plane cut for the SHG intensity, measured in kW/cm$^2$. Panels (e,f): temperature profile, showing the thermal gradient in the nonlinear crystal interaction.}
    \label{fig:ex2}
\end{figure*}

\noindent
To illustrate the most general utility of this package, we solve the numerical algorithm shown in Figure~\ref{fig:flowchart}, in which the heat equation (Eqs. \ref{eq:heatEqSS}-\ref{eq:convection}) and the SSFM (Eq.~\ref{eq:CWEsmatrix}) are solved systematically, until the overall system reaches the steady state. A typical experimental procedure to find the maximum efficiency, for a given input pump power, a fixed beam waist and a fixed focal position, is to scan the crystal temperature. Figure~\ref{fig:ex2}(a) shows this scanning procedure for different pumping levels for a pump beam waist of $w_{\mathrm{F}}=29~\upmu$m, focused at the center of a $2\times 1\times 30$~mm crystal (width, height, length). The inset shows the maximum efficiency (peaks of the curves) as a function of temperature. The conversion efficiency as well as the generated SH power are shown in panel (b); these results are in a excellent agreement with the measurements performed in, e.g., Ref.~\citep{kumar2011high}. In panels (c) and (d), the fundamental and SH electric fields propagation are shown in a $yz$-plane cut. In panel (e) and (f), the temperature profile is shown, where the thermal lensing effect clearly shifts the focal position shown in panel (d). 
The compilation of this example is detailed in Listing~\ref{lst:compilation}, case \verb|#2|, where the flag \verb|-DTHERMAL| must be included. 

\section{Conclusions}
\label{sec:conclusions}
\noindent
In this work, a package written in the object-oriented programming paradigm in the \cpp/CUDA language that calculates the efficiency of SHG process in the absence and presence of thermal effects has been presented. Our package implements coupled differential equations that describe second-order nonlinear interactions in a dielectric crystal. Our model includes the effect of linear and nonlinear diffraction and absorption, which contribute to thermal lensing effect. The latter can lead to an undesirable operating regime in which the SHG efficiency degrades due to longitudinal and transverse temperature gradients in the nonlinear crystal. The performance of the package has been measured in terms of its speedup, by comparing the CPU and GPU implementations in completely analogous codes. Speedups of between 17x and 43x were obtained for calculations that do not include the temperature profiles, while with the inclusion of thermal effects we measured speedups of between 4x and 13x.

While the examples shown in our package used a quasi-phase-matched nonlinear crystal of MgO:sPPLT, with minimal modifications to the corresponding header file users can change the medium to any other material, including birefringently phase-matched crystals. In such a case, the code also includes the corresponding spatial walk-off angle term. 

Finally, the package provides a useful computational tool which can be exploited by scientists for prediction of SHG conversion efficiency and design optimisation prior to practical system implementation. Likewise, the package can be useful for those users in the early stages of study in the subject who require numerical implementation of the SHG process. This package can also serve as a starting point for other similar three-wave mixing processes.

\section*{Declaration of competing interest}
The authors declare that they have no known competing financial interests or personal relationships that could have appeared to influence the work reported in this paper.

\section*{Data availability}
Data will be made available on request.

\section*{Acknowledgments}
We gratefully acknowledge funding from the Ministerio de Ciencia e Innovaci\'on (MCIN) and the State Research Agency (AEI), Spain (Project Nutech PID2020-112700RB-I00); Project Ultrawave EUR2022-134051 funded by MCIN/AEI and by the European Union NextGenerationEU/PRTR; Severo Ochoa Programme for Centres of Excellence in R\&D (CEX2019-000910-S); Generalitat de Catalunya (CERCA); Fundaci\'on Cellex; Fundaci\'o Mir-Puig. S. Chaitanya Kumar acknowledges support of the Department of Atomic Energy, Government of India, under Project Identification No. RTI 4007.
The authors would also like to express their sincere gratitude to Maximiliano Gilberto for his insightful discussions about several topics this work includes.

\appendix

\section{Diffraction term} \label{asec:diffraction}
The diffraction terms in Eq.~\ref{eq:CWEsmatrix}, $\nabla^2_{\perp}$, are solved separately in the momentum space by performing 2DFT. The basic equation to solve is
\begin{equation}\label{eq:eqA1}
    \derparvar{A_j}{z} =-\frac{i}{2k_j}\left(\frac{\partial^2}{\partial x^2}+\frac{\partial^2}{\partial y^2}\right) A_j
\end{equation}
where $j\in\{\mathrm{F},\mathrm{SH}\}$ labels each of the electric fields. The solution to the Eq.~\ref{eq:eqA1} is obtained by taking the 2DFT
\begin{equation}\label{eq:eqA2}
    A_j(x,y,z) = \int\displaylimits_{-\infty}^{+\infty}\int\displaylimits_{-\infty}^{+\infty} \tilde{A}_j(q_x,q_y,z) e^{i 2\pi(q_x x+ q_y y)} dq_x dq_y,
\end{equation}
which leads to
\begin{equation}\label{eq:eqA3}
    \derparvar{\tilde{A}_j}{z} = i\frac{2\pi^2}{k_j}\left(q_x^2 + q_y^2\right)\tilde{A}_j,
\end{equation}
where $q_x$ and $q_y$ are the spacial frequencies in the transverse directions. Equation~\ref{eq:eqA3} is solved numerically as
\begin{equation}\label{eq:eqA4}
    \tilde{A}_j(q_x,q_y,z+\Delta z) =\tilde{A}_j(q_x,q_y,z)e^{\hat{Q}_j\Delta z},
\end{equation}
where 
\begin{equation}\label{eq:eqA5}
    \hat{Q}_j = \frac{2i\pi^2}{k_j}\left(q_x^2 + q_y^2\right)
\end{equation}
is the propagator of the electric field $j$.

\section{Package classes and their methods} \label{asec:classes}
In the following, we present the package classes with their relevant methods. The definition of any package class is summarized in Listing~\ref{lst:generic_class}. As can be seen, the data members are pointers that point to memory addresses where quantities such as temperature (real) or electric fields (complex) are stored. Both constructor and destructor are defined in \textit{host} and \textit{device}. Here, \verb|functions()| represents any method included in a generic class, \verb|ClassName{}|. In turn, the methods invoke CUDA kernels that perform some operation on data members belonging to their own or other classes.
\begin{lstlisting}[caption={Basic structure of the package classes},label={lst:generic_class},language=C++, basicstyle=\ttfamily,
  showstringspaces=false,
  commentstyle=\color{gray!50!black},
  keywordstyle=\color{blue},
  basicstyle=\ttfamily\footnotesize,] 
class ClassName{ // Difine the class
public:
    // Data members
    real_t *rMat; complex_t *clxMat;
    __host__ __device__
    ClassName(){ // Constructor
        cudaMalloc((void **)&rMat,SIZE*sizeof(real_t)); 
        cudaMalloc((void **)&cpxMat,SIZE*sizeof(complex_t));
    }
    __host__ __device__
    ~ClassName(){ // Destructor
        cudaFree((void *)rMat);
        cudaFree((void *)clxMat);
    }
    functions() // Methods to access data members
}
\end{lstlisting}

\subsection{Efields class} \label{asec:efieldsclass}
The \verb|Efields| class has seven data members representing the electric fields in both coordinate (\verb|Pump| and \verb|Signal|) and momentum (\verb|PumpQ| and \verb|SignalQ|) space as well as the field propagators (\verb|PropPump| and \verb|PropSignal|, see \ref{asec:diffraction}, Eq.~\ref{eq:eqA5}). 
The basic definition of this class is summarized in Listing~\ref{lst:efields}.
\begin{lstlisting}[caption={Class Efields},label={lst:efields},language=C++, basicstyle=\ttfamily,
  showstringspaces=false,
  commentstyle=\color{gray!50!black},
  keywordstyle=\color{blue},
  basicstyle=\ttfamily\footnotesize,]
class Efields{
public:
    // Seven data members
    complex_t *Pump, *Signal, *PumpQ, *SignalQ;
    complex_t *PropPump, *PropSignal, *AuxQ;
  
    __host__ __device__ Efields(){...} // Constructor
    __host__ __device__~Efields(){...} // Destructor

    // Methods to access data members
    void setInputPump(<PARAMETERS>)
    void setNoisyField()
    void setPropagators(<PARAMETERS>)
}
\end{lstlisting}
The method \verb|setInputPump(<PARAMETERS>)| sets the pump electric field, while \verb|setNoisyField()| fills the SH electric field matrix with complex random numbers. The method \verb|setPropagators(<PARAMETERS>)| sets the beam propagators for fundamental and SH electric fields. This function also includes the walk-off angle term, $\tan(\rho)\partial A/\partial x$, although we set $\rho=0$ in \verb|Crystal.h|.

\subsection{Tfield class} \label{asec:tfieldclass}
The \verb|Tfield| class has four data members representing the temperature field, $T(\Vec{r})$, and the internal heat source, $\dot{q}(\Vec{r})$. The basic definition of this class is summarized in Listing~\ref{lst:tfield}.
\begin{lstlisting}[caption={Class Tfield},label={lst:tfield},language=C++, basicstyle=\ttfamily,
  showstringspaces=false,
  commentstyle=\color{gray!50!black},
  keywordstyle=\color{blue},
  basicstyle=\ttfamily\footnotesize,]
class Tfield{
public:
    // four data members
    real_t *Tinic, *Tfinal, *Taux, *Q;
    
    __host__ __device__ Tfield(){...} // Constructor
    __host__ __device__~Tfield(){...} // Destructor

    // Methods to access data members
    void setTemperature(<PARAMETERS>)
    void setBottomOvens(<PARAMETERS>)
    void setOvenSurrounded(<PARAMETERS>)
    void upDate(<PARAMETERS>)
    real_t checkConvergence()
    void setInitialQ()
    void upDateQ(<PARAMETERS>)    
}
\end{lstlisting}
The method \verb|setTemperature(<PARAMETERS>)| initializes the crystal temperature. \verb|setBottomOvens(<PARAMETERS>)| and \verb|setOvenSurrounded(<PARAMETERS>)| set the oven temperature for an oven configuration corresponding to a single-bottom oven or an oven surrounding the nonlinear crystal. The methods \verb|upDate(<PARAMETERS>)| (computes the finite-elements method in Eq.~\ref{eq:heatEqSSdisc}) and \verb|upDateQ(<PARAMETERS>)| update the temperature field and the internal heat source, respectively. \verb|setInitialQ()| initializes the internal heat source, $\dot{q}(\Vec{r}) = 0$. Finally, the method \verb|checkConvergence()| compares the temperature field between two consecutive iterations and decides whether or not the system has reached the steady state.

\subsection{PhaseMatching class} \label{asec:pmclass}
The \verb|PhaseMatching| class has two data members representing the mismatch factor, $ \dk(\Vec{r})$. The basic definition of this class is summarized in Listing~\ref{lst:pm}.
\begin{lstlisting}[caption={Class PhaseMatching},label={lst:pm},language=C++, basicstyle=\ttfamily,
  showstringspaces=false,
  commentstyle=\color{gray!50!black},
  keywordstyle=\color{blue},
  basicstyle=\ttfamily\footnotesize,]
class PhaseMatching{
public:
    real_t *DK, *DKint; // two data members
  
    __host__ __device__ PhaseMatching(){...} // Constructor
    __host__ __device__~PhaseMatching(){...} // Destructor

    // Methods to access data members
    void IntegrateDK(<PARAMETERS>)
    void setDKFromTemperature(<PARAMETERS>)
    void setInicialDKConstant(<PARAMETERS>)    
}
\end{lstlisting}
The method \verb|IntegrateDK(<PARAMETERS>)| performs the integral of Eq.~\ref{eq:integralpm}. \verb|setDKFromTemperature(<PARAMETERS>)| updates the mismatch factor in the coordinates after thermal calculations. \verb|setInicialDKConstant(<PARAMETERS>)| fills the mismatch factor matrix with the corresponding phase-matching calculated value for simulations without thermal evolution.

\subsection{Solver class} \label{asec:solverclass}
The \verb|PhaseMatching| class has ten data members to perform the fourth-order Runge-Kutta (RK4) method for solving the CWEs. The basic definition of this class is summarized in Listing~\ref{lst:solver}.
\begin{lstlisting}[caption={Class Solver},label={lst:solver},language=C++, basicstyle=\ttfamily,
  showstringspaces=false,
  commentstyle=\color{gray!50!black},
  keywordstyle=\color{blue},
  basicstyle=\ttfamily\footnotesize,]
class Solver{
public:
    real_t *ks,...; // ten data members
  
    __host__ __device__ Solver(){...} // Constructor
    __host__ __device__~Solver(){...} // Destructor

    // Methods to access data members
    void diffraction(<PARAMETERS>);
    void solverRK4(<PARAMETERS>);
    void SSFM(<PARAMETERS>);
    void CWES(<PARAMETERS>);
    void run(<PARAMETERS>);    
}
\end{lstlisting}
The method \verb|diffraction(<PARAMETERS>)| computes the diffraction term for fundamental and SH electric fields shown in Eq.~\ref{eq:eqA4}. \verb|solverRK4(<PARAMETERS>)| performs the RK4 method. \verb|SSFM(<PARAMETERS>)| computes the Split-Step Fourier method, and \verb|CWES(<PARAMETERS>)| applies the SSFM along the crystal propagation, solving the nonlinear system in Eq.~\ref{eq:CWEsmatrix}. Finally, the method \verb|run(<PARAMETERS>)| executes the full model and is called in the main file, receiving the experimental parameters required to perform the simulation (see Model execution in Listing~\ref{lst:mainfile}).

\bibliographystyle{elsarticle-num}
\bibliography{main}

\end{document}